\documentclass[prl]{revtex4}

\usepackage{graphicx}
\usepackage{dcolumn}
\usepackage{bm}

\begin{document}
\renewcommand{\theequation}{\arabic{equation}}

\title{Maximal acceleration and radiative processes\\}

\author{Giorgio Papini}
\altaffiliation[Electronic address:]{papini@uregina.ca}
\affiliation{Department of Physics and Prairie Particle Physics
Institute, University of Regina, Regina, Sask, S4S 0A2, Canada}


\begin{abstract}

We derive the radiation characteristics of an accelerated, charged particle in a model due to Caianiello in which the proper
acceleration of a particle of mass $m$ has the upper limit $\mathcal{A}_m=2mc^3/\hbar$. We find two power laws, one applicable
to lower accelerations, the other more suitable for accelerations closer to $\mathcal{A}_m$ and to the related physical
singularity in the Ricci scalar. Geometrical constraints and power spectra are also discussed.
By comparing the power laws due to the maximal acceleration with that for particles in gravitational fields, we find that
the model of Caianiello allows, in principle, the use of charged particles as tools to distinguish inertial from gravitational fields locally.

\end{abstract}

 \maketitle

\setcounter{equation}{0}
\section{Introduction}
In a model aimed at providing quantum mechanics
with a geometrical framework \cite{cai3.1,cai3.2,cai3.3}, Caianiello introduced the upper limit ${\cal A}_m =
2mc^{3}/\hbar$ to the proper acceleration of a particle. This mass dependent limit, or maximal
acceleration (MA), can be derived from quantum mechanical
considerations and the fact that the acceleration is largest in
the particle rest frame \cite{ca,pw,papinix}. The absolute value
of the proper acceleration therefore satisfies the inequality $ a
\leq {\cal A}_m $. No counterexamples are known to the validity of
this inequality.

Classical and quantum arguments supporting the existence of a MA
have been given in the literature \cite{prove.1,prove.2,prove.3,prove.4,prove.5,prove.6,prove.7,prove.8,prove.9,prove.10,prove.11,prove.11bis,prove.12,prove.13,
prove.14,prove.15,
prove.16,prove.17,prove.18,prove.19,prove.20,prove.21,
prove.22,prove.23,prove.24,prove.25,prove.26,wh,b.1,b.2}. MA also
appears in the context of Weyl space \cite{pap.1,pap.2,pap.3,pap.4} and of a
geometrical analogue of Vigier's stochastic theory \cite{jv}. It is invoked as a tool to rid
black hole entropy of ultraviolet divergences \cite{McG} and is
at times regarded as a regularization procedure \cite{nesterenko}
that avoids the introduction of a fundamental length \cite{gs},
thus preserving the continuity of space-time.

An upper limit on the acceleration also exists in string theory \cite{gsv.1,gsv.2,gsv.3,gasp.1,gasp.2} when the
acceleration induced by the background gravitational field reaches
the critical value $a_c = \lambda^{-1} = (m\alpha)^{-1}$ where
$\lambda$, $m$ and $\alpha^{-1}$ are string size, mass and
tension. At accelerations larger than $a_c$ the string extremities
become casually disconnected. Frolov and Sanchez \cite{fs} have
found that a universal critical acceleration must be a general
property of strings. It is the same cut--off required by Sanchez
in order to regularize the entropy and the free energy of quantum
strings \cite{sa2}.

Applications of Caianiello's model include cosmology \cite{infl.1,infl.2,infl.3},
the dynamics of accelerated strings \cite{Feo.1,Feo.2} and neutrino
oscillations \cite{8,qua.1,qua.2} and the determination of a lower
neutrino mass bound \cite{neutrinobound}. The model also makes the
metric observer--dependent, as conjectured by Gibbons and Hawking
\cite{Haw}.

The model has been applied to classical \cite{sch} and quantum particles \cite{boson} falling in the
gravitational field of a collapsing, spherically symmetric object described
by the Schwarzschild metric
and also to the Reissner-Nordstr\"om \cite{reiss} and Kerr \cite{kerr} metrics.
In the model, the end product of stellar collapse is
represented by compact, impenetrable astrophysical objects whose
radiation characteristics are similar to those of known bursters
\cite{papiniz}.

The consequences of the model for the classical electrodynamics of
a particle \cite{cla}, the mass of the Higgs boson \cite{Higgs.1,Higgs.2}
and the Lamb shift in hydrogenic atoms \cite{lamb} have been
worked out.
MA effects in muonic
atoms \cite{muo} and on
helicity and chirality of particles \cite{chen} have also been investigated.

Most recently Rovelli and Vidotto in two important works have found evidence for MA
and singularity resolution in covariant loop quantum gravity \cite{rovelli1},\cite{rovelli2}.

Caianiello's model is based on an embedding procedure \cite{sch}
that stipulates that the line element experienced by an
accelerating particle is represented by
\begin{equation} \label{eq1}
d\tau^2=\left(1+\frac{g_{\mu\nu}\ddot{x}^{\mu}\ddot{x}^{\nu}}{{\cal
A}_m^2}
\right)g_{\alpha\beta}dx^{\alpha}dx^{\beta}=\left(1+\frac{a^2(x)}{{\cal
A}_m^2}
 \right) ds^2\equiv
\sigma^2(x) ds^2\,,
\end{equation}
where $g_{\alpha\beta}$ is a background gravitational field. The
effective space-time geometry given by (\ref{eq1}) therefore
exhibits mass-dependent corrections that in general induce
curvature and violations of the equivalence principle.

\begin{figure}
\centering
\includegraphics[width=0.3\textwidth]{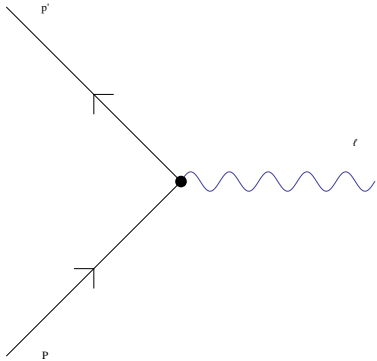}
\caption{\label{fig:Feynman2}This process, normally forbidden
by conservation of energy and momentum, is allowed in the MA
model.}
\end{figure}

The purpose of this paper is to determine the radiation characteristics of
an accelerating charged fermion in the field of the metric (\ref{eq1}) in
the process
described by Fig.\ref{fig:Feynman2}. We tackle the problem in two distinct ways,
first by using
solutions of the covariant Dirac equation that are exact to first order in the
metric deviation $\gamma_{\mu\nu}=g_{\mu\nu}-\eta_{\mu\nu}$, where
$\eta_{\mu\nu}$ is the metric of flat space-time (of signature -2).
In these relativistic solutions
the MA field of (\ref{eq1}) appears in a phase factor and the Riemann tensor is linearized.
This approach is suitable to study the effects of MA when $ a
\ll {\cal A}_m $. In the second approach, more suitable for extreme accelerations, we consider the effects of the
Ricci scalar and of its physical singularity at $a=\mathcal{A}_{m}$ on the index of refraction of the incoming fermion. In this way some aspects of the
problem at more elevated accelerations can be considered.
The single-vertex diagram selected represents, potentially, the largest contribution to the radiation process. It is however
forbidden kinematically in Minkowski space. Let us consider in fact the process
\begin{equation}\label{2}
  P_\mu=p_\mu^\prime + \ell_\mu\,,
\end{equation}
in which an incoming massive particle of
momentum $P_{\mu}$ and dispersion relation $P_{\mu}P^{\mu}
=m_{1}^2$ produces a photon of momentum $\ell_{\mu}\ell^{\mu}=0$,
while the outgoing particle has momentum $p'_{\mu}p'^{\mu}=m^2$.
Conservation of energy-momentum requires
$P_{\mu}=p'_{\mu}+\ell_{\mu}$. In the rest frame of $P$ we have
$0=\vec{p'}+\vec{\ell}$, which gives $\vec{\ell}=-\vec{p'}$,
$P_{0}=m_{1}$ and again $m_{1}=\sqrt{\ell^2+m^2}+\ell$. Then
$(m_{1}-\ell)^2=\ell^2 +m^2$ leads to $\ell=(m_{1}^2-m^2)/2m_{1}$
which shows that for $m_{1}=m$, the case considered, we get $\ell=0$ and the process
becomes physically meaningless. There are processes, however, in
which massive particles emitting a photon are not kinematically
forbidden. This is certainly the case when
gravitation alters the dispersion relations of at least one of the
particles involved \cite{PRD82,GAL}. This is also the case with MA \cite{MPLA}.

\section{the Dirac equation and MA}
Because MA acts as a gravitational field according to (\ref{eq1}), we can also expect
changes in the index of refraction and dispersion relations of the incoming,
accelerating fermion.

Let us
consider, for simplicity, the case of hyperbolic motion in which
the fermion always experiences a constant acceleration $a_\mu
a^\mu = 1/\xi^2$. From (\ref{eq1}) we immediately obtain the
modified Minkowski metric
\begin{equation}\label{3}
  d\tau^2=\left(1-\frac{1}{{\cal A}_m^2 \xi^2}\right)(dt^{2}-
  dx^{2}-dy^{2}-dz^2)\,,
\end{equation}
for that region of space-time delimited by the physical branch of
the hyperbola $t^2 -z^2\equiv\xi^{2}$.
The behavior of spin-1/2 particles in the presence of a
gravitational field $g_{\mu\nu}$ is determined by the covariant
Dirac equation
\begin{equation}\label{DiracEquation}
  [i\gamma^\mu(x){\cal D}_\mu-m]\Psi(x)=0\,,
\end{equation}
where ${\cal D}_\mu=\nabla_\mu+i\Gamma_\mu (x)$, $\nabla_\mu$ is
the covariant derivative, $\Gamma_{\mu}(x)$ the spin connection
and the matrices $\gamma^{\mu}(x)$ satisfy the anti-commutation relations
$\{\gamma^\mu(x), \gamma^\nu(x)\}=2g^{\mu\nu}(x)$. Both
$\Gamma_\mu(x)$ and $\gamma^\mu(x)$ can be related to the usual
constant Dirac matrices $\gamma^{\hat{\alpha}}$ by using the tetrad fields $e_{\hat
\alpha}^\mu$ and the relations
\begin{equation}\label{II.2}
  \gamma^\mu(x)=e^\mu_{\hat \alpha}(x) \gamma^{\hat
  \alpha}\,,\qquad
  \Gamma_\mu(x)=-\frac{1}{4} \sigma^{{\hat \alpha}{\hat \beta}}
  e^\nu_{\hat \alpha}e_{\nu\hat{\beta};\, \mu}\,,
\end{equation}
where $\sigma^{{\hat \alpha}{\hat \beta}}=\frac{i}{2}[\gamma^{\hat
\alpha}, \gamma^{\hat \beta}]$. A semicolon and a comma are
frequently used as alternative ways to indicate covariant and
partial derivatives respectively.
Then the connection is represented
by
\begin{equation}
\Gamma_{\mu}=\frac{1}{2}\sigma^{\hat{\alpha}\hat{\beta}}\omega_{\mu \hat{\alpha}\hat{\beta}}\,{,}
\end{equation}
where
\begin{equation}
\sigma^{\hat{\alpha}\hat{\beta}}=\frac{1}{4}\,[\gamma^{\hat{\alpha}},\gamma^{\hat{\beta}}],\quad
\omega_{\mu\,\,\,\,\hat{\beta}}^{\,\,\,\,\hat{\alpha}}=
(\Gamma^{\lambda}_{\mu\nu}e_{\lambda}^{\,\,\,\,\hat{\alpha}}-
\partial_{\mu}e_{\nu}^{\,\,\,\,\hat{\alpha}})e^{\nu}_{\,\,\,\,\hat{\beta}}
\end{equation}
\begin{equation}
\Gamma^{\lambda}_{\mu\nu}=\frac{1}{2}
g^{\lambda\alpha}(g_{\alpha\mu,\nu}+g_{\alpha\nu,\mu}-g_{\mu\nu,\alpha}),\quad
\gamma^{\mu}(x)=e^{\mu}_{\,\,\,\,\hat{\alpha}}\gamma^{\hat{\alpha}},\quad
e^{\mu}_{\,\,\,\,\hat{\alpha}}e_{\mu}^{\,\,\,\,\hat{\beta}}
=\delta_{\hat{\alpha}}^{\,\,\,\,\hat{\beta}}\,{.}
\end{equation}
The form of (1) determines the tetrad field and the connection. From (3)-(8) we find
\begin{equation}
e_{\mu}^{\,\,\,\,\hat{\alpha}}(x)=\sigma(x)\delta_{\mu}^{\,\,\,\,\hat{\alpha}},\quad
\Gamma_{\mu}=\sigma^{\hat{\alpha}\hat{\beta}}\eta_{\hat{\alpha}\mu}(\ln\sigma)_{,\hat{\beta}}\,{.}
\end{equation}
The solution of (\ref{DiracEquation}) exact to first order in $\gamma_{\mu\nu}=f\eta_{\mu\nu}$
where $f=\sigma^{2}-1$, is \cite{caipap,singh,pap1}
\begin{equation}\label{PsiSolution2}
  \Psi(x)=-\frac{1}{2m}\left(-i\gamma^\mu(x){\cal
  D}_\mu-m\right)e^{-i\Phi_T}\Psi_0(x)\,,
\end{equation}
where $\Phi_T=\Phi_s+\Phi_G$ is of first order in
$\gamma_{\alpha\beta}(x)$. The factor $ -1/2m $ on the r.h.s. of
(\ref{PsiSolution2}) is required by the condition that both sides
of the equation agree when the MA contribution vanishes. Similar solutions can also be found for all covariant wave equations
\cite{papb,pap2,pap3,pap4}. We choose
$\Psi_{0}(x)\propto e^{-ip_{\alpha}x^{\alpha}}$ and drop carets in what follows.
In (\ref{PsiSolution2}) $\Phi_{S}(x)=\mathcal{P}\int_{P}^x dz^\lambda \Gamma_\lambda (z)$
and
\begin{equation}\label{PhiGtilde}
  \Phi_G(x)=-\frac{1}{4}\int_P^xdz^\lambda\left[\gamma_{\alpha\lambda,
  \beta}(z)-\gamma_{\beta\lambda,
  \alpha}(z)\right]((x^{\alpha}-z^{\alpha})p^{\beta}-(x^{\beta}-z^{\beta})p^{\alpha})+
  \frac{1}{2}\int_P^x dz^\lambda\gamma_{\alpha\lambda}p^\alpha\,.
\end{equation}
It follows from (\ref{DiracEquation}) and (\ref{PhiGtilde}) that the physical
momentum of the incoming fermion is
\begin{equation}\label{A2}
  P_{\mu} =
  -p_{\mu} - \Phi_{G, \mu}=
   -p_{\mu}-\frac{1}{2}\,\gamma_{\alpha \mu}(x)p^\alpha
   +\frac{1}{2}\, \int_P^x
   dz^\lambda(\gamma_{\mu\lambda,\beta}(z)-\gamma_{\beta\lambda,\mu}(z))p^\beta\,.
\end{equation}
For hyperbolic motion in the $(t,z)$-plane we find
$P_{3}=-p_{3}-\Phi_{G,3}=-p_{3}-\frac{p_{0}^{2}f}{2p}$, where $p\equiv p^3$.
It therefore follows that the particle's momentum remains finite
even at the MA limit $f=-1$, and so does the index of refraction
\begin{equation}\label{N}
N_{f}=\frac{|\vec{P}|}{P_{0}}=\frac{y}{\sqrt{1+y^2}}\left\{\frac{y^2-\frac{f}{2}(1+y^2)}{y^2+\frac{f}{2}(1+y^2)}\right\}\,,
\end{equation}
where $y\equiv p/m$. It also follows that $N_{f}\sim 1$ for $ f=0$ and that there are no allowable values
of $f$ for which $N_f$ vanishes. In addition $N_f $ diverges for values of $y$ and $f$ such that $y^2=-f/2/(1+f/2)$.
All these results also apply to self-accelerating particles. Though obtained
by linearizing (\ref{DiracEquation}), solution (\ref{PsiSolution2})
can be extended to any order in $\gamma_{\mu\nu}=f\eta_{\mu\nu}$.

\section{The power emitted}
It now is possible to calculate the power radiated as photons in the process
of Fig.\ref{fig:Feynman2} following the procedure outlined in \cite{PRD82,GAL}.
We find
\begin{equation}\label{W}
W=\frac{1}{8(2\pi)^2}\int\delta^{4}(P-p'-l)\frac{|M|^2}{P^3}\Theta(p_{0}')\delta(p'^{2}-m^2)d^{4}p'd^{3}\ell\,,
\end{equation}
where
\begin{equation}\label{W1}
|M|^2=Z^2e^2\left|\left[-4(p'_{\alpha}P^{\alpha})+8(p_{\alpha}P^{\alpha})\right]+2\left[2\gamma_{\alpha\beta}p'^{\alpha}p^{\beta}+
\gamma_{\alpha\beta}\eta^{\alpha\beta}(-p'_{\lambda}p^{\lambda})+m^2)\right]\right|\,,
\end{equation}
and, neglecting the spin-MA interaction, $P_{\alpha}=p_{\alpha}+\Phi_{G,\alpha}$.
Using the identity $\int \frac{d^3p'}{2p'_0}=\int d^4p'\Theta(p'_{0})\delta(p'^2 -m^2)$, integrating over $d^{4}p'$
and writing the on-shell condition for $p'$ as
\begin{equation}\label{cos}
\delta(\cos\theta-\frac{m^2-P_{\alpha}P^{\alpha}+2\ell_{0}P_{0}}{2|\vec{P}||\vec{\ell}|})\,,
\end{equation}
we obtain
\begin{equation}\label{W2}
W= \frac{(Ze)^2}{8\pi}\int d\ell \frac{\ell}{P|\vec{P}|}\left|\frac{f-1}{2}(P_{\alpha}P^{\alpha})-\frac{m^2}{2}(f+1)+(f+2)(p_{\alpha}P^{\alpha})\right|\,,
\end{equation}
where $P^{3}\equiv P$. Using (\ref{A2}) to calculate the components of $P_{\mu}$ and substituting the results in (\ref{W2}) we obtain
\begin{equation}\label{W3}
W\approx \frac{5(Ze)^2}{32\pi}\frac{m^2\ell^2}{p^2}\left|f\left(1+\frac{fp_{0}^2}{p^2}\right)\right|=\frac{5(Ze)^2}{32\pi}\frac{\ell^2 }{y^2}\left|f+\left(1+\frac{1}{y^2}\right)f^2\right|\,.
\end{equation}
Note that (\ref{W3}) vanishes as $f\rightarrow 0$ (vanishing acceleration) and diverges as $y\rightarrow 0$. This infrared divergence
has already been discussed in \cite{PRD82} and can be removed \cite{weinberg}.
The power spectrum follows immediately from (\ref{W3}) and is
\begin{equation}\label {spectrum}
\frac{dW}{d\ell}= \frac{5 (Ze)^2}{16\pi}\frac{\ell }{y^2}\left|f+\left(1+\frac{1}{y^2}\right)f^2\right|\,.
\end{equation}
These results require that the inequalities $-1\leq \cos\theta \leq 1$ be satisfied. We find that they
can be both satisfied if
\begin{equation}\label{ineq}
\frac{m^2 f}{2p_{0} +2p+\frac{m^2 f(m^2 +2p^2)}{p^2 p_{0}}}\leq \ell\leq\frac{m^2 f}{2p_{0} -2p+\frac{m^2 f(m^2 +2p^2)}{p^2 p_{0}}}\,.
\end{equation}
In the high momentum approximation $p\gg m$, (\ref{ineq}) becomes
$0\leq\ell\leq pf/(1+3f/2)$. The condition that $\ell>0$ requires in addition that $f<-2/3$. By substituting $f\sim -1$ and $\ell\sim 2p$ in the expression for
$\cos\theta$ we find $\cos\theta\sim 1$ and radiation is in the forward direction. At
the other extreme, $\ell \sim 0$, we also find $\theta \sim 0$. In the approximation used, radiation is therefore possible
in the interval of accelerations $-1< f < -2/3$.

\section{The effect of the curvature scalar}

A better perspective on the behaviour of a fermion closer to the limit $\mathcal{A}_m$ can
be acquired by introducing a different
approximation to the dispersion relation of the fermion in the MA field.
Let us therefore multiply (\ref{DiracEquation}) on the left with
$(-i {\gamma}^\nu (x) D_\nu  - {mc}/{\hbar}),$\,.
We obtain the second-order equation
\begin{equation}
\left(\gamma^\mu(x) \gamma^\nu(x) D_\mu D_\nu + \frac{m^2 c^2}{\hbar^2}\right)
{\psi}(x) \ = \ 0\,,
\label{ins1}
\end{equation}
which, on using the relations $[D_\mu, D_\nu] \ = \ -\frac{i}{4}
\sigma^{\alpha \beta}R_{\alpha \beta \mu \nu}$ and $\sigma^{\mu \nu} \sigma^{\alpha \beta}R_{\mu \nu \alpha \beta} \ = \ 2R$, reduces to
\begin{equation}\label{KG}
\left(g^{\mu \nu} D_\mu D_\nu - \frac{R}{4} + \frac{m^2 c^2}{\hbar^2}\right)
{\psi}(x) \ = \ 0\,.
\label{ins2}
\end{equation}
In the eikonal approximation $\nabla_\mu \psi=ip_\mu \psi$, (\ref{KG}) yields
\begin{equation}\label{5}
  -g^{\mu\nu}p_\mu p_\nu -\frac{R}{4}+m^2=0\,.
\end{equation}
From this relation one obtains the index of refraction
\begin{equation}\label{nf}
 n_f = \frac{|{\bf
 p}|}{p_0}=\sqrt{g^{00}}\sqrt{1-\frac{m^2-R/4}{g^{00}p_0^2}}\,,
\end{equation}
which contains the scalar curvature $R$
and its physical singularity at $ f=-1 $ and therefore differs substantially from (\ref{N}). In (\ref{nf}) $|{\bf p}|=\sqrt{|g^{ij}|p_i p_j}$ and the radiation angle
between ${\bf p}$ and ${\bf \ell}$ momenta is given by,
\begin{equation}\label{7}
  \cos \theta'=\frac{g^{ij}p_i \ell_j}{ \sqrt{g^{ij} p_i p_j} \sqrt{\eta^{ij}
  \ell_i \ell_j}}\,.
\end{equation}

Multiplying
(\ref{2}) by $p_\mu$ and $\ell_\mu$, and using the dispersion
relations (\ref{5}) and $\ell_\mu \ell^\mu=0$, we get
\begin{equation}\label{a}
  m^2-\frac{R}{4}=g^{\mu\nu}p_\mu p_\nu'+g^{\mu\nu} p_\mu \ell_\nu\,,\quad g^{\mu\nu} p_\mu \ell_\nu=
  g^{\mu\nu}p_\mu' \ell_\nu\,.
\end{equation}
On the other hand, assuming that after the emission of the photon,
the fermion propagates with vanishing four-acceleration, the
multiplication of (\ref{2}) by $p_\mu^{\prime}$ and the dispersion
relation $\eta^{\mu\nu}p_\mu^\prime p_\nu^\prime=m^2$ give
\begin{equation}\label{b}
  g^{\mu\nu} p_\mu' p_\nu=m^2 +g^{\mu\nu}p_\mu' \ell_\nu\,.
\end{equation}
Equations (\ref{a}) and (\ref{b}) lead to the result
\begin{equation}\label{c}
 g^{\mu\nu} p_\mu \ell_\nu = -\frac{R}{8}\,,\quad
  R=\frac{6{\cal A}_m^2 f^3}{(1+f)^3}\,.
 \end{equation}
The angle (\ref{7}) therefore becomes
\begin{equation}\label{8}
  \cos \theta' =\frac{-g^{00}-\displaystyle{\frac{R}{8\ell_0 p_0}}}{n_f \sqrt{g^{00}}}\,.
\end{equation}
It is convenient, in view of the following discussion, to write the
explicit expression for the index of refraction (\ref{nf})
\begin{equation}\label{nf1}
 n_f=\frac{1}{\sqrt{1+f}}\sqrt{1-\frac{1+f}{y^2}\left(1 + \frac{6}{(1+f)^3}\right)}\,,
\end{equation}
which gives $n_f =\sqrt{1-(1+3/2)/y^2}\sim 1$ for $f=0$ and high values of $y$.
We also find that $n_f$ is always positive for
\begin{equation}\label{real}
f\geq \frac{1}{3} (-3 + y^2) + \frac{y^4}{3 (-81 + y^6 + 9 \sqrt{81 - 2 y^6]})^{\frac{1}{3}}}+
 \frac{1}{(-81 + y^6 + 9 \sqrt{81 - 2 y^6})^{\frac{1}{3}}}\,.
\end{equation}
There also are two complex conjugate solutions of $n_f =0$ that suggest complex optical properties for
MA space-time. We will not investigate them in this work.

\begin{figure}
\centering
\includegraphics[width=0.3\textwidth]{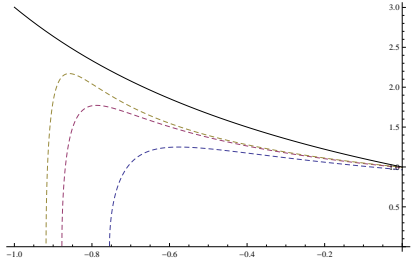}
\caption{\label{fig:QGEOM}The dotted curves refer to $n_f$ and, respectively, to $y=30, 20, 10$ from the top down.
The upper curve refers to $N_{f}$ for $k=30$ and is not particularly sensitive to the values of $y$ in the range considered.}
\end{figure}

A comparison of the general behaviour of $n_f$ and $N_f$ is given in Fig. \ref{fig:QGEOM}.

The calculation of the diagram of Fig.\ref{fig:Feynman2} now proceeds as for (\ref{W}) and (\ref{W1}), where $P_3$ is now replaced by $p_3$, and use is made of
(\ref{5})-(\ref{nf1}). We obtain
\begin{equation}\label{WW}
W'=\frac{(Ze)^2 \ell^2}{32\pi y^2}\left|\frac{y\sqrt{1+f}}{\sqrt{(1+f)(1+y^2)+\frac{6 f^3}{(1+f)^3}-1}}-1\right|\,,
\end{equation}
which vanishes for $f=0$ and has singularities for $f$ and $y$ satisfying the equation
$(1+f)(1+y^2) +6f^3/(1+f)^3 =1$. The infrared divergence for $y\rightarrow 0$ can also be removed \cite{weinberg}.
The power spectrum $\frac{dW'}{d\ell}$ follows from (\ref{WW}).

\begin{figure}
\centering
\includegraphics[width=0.3\textwidth]{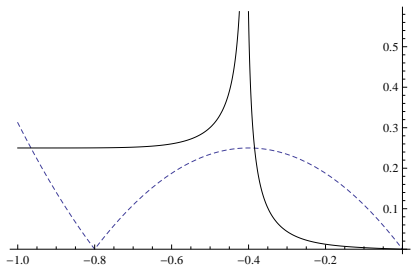}
\caption{\label{fig:QGEOM1}The dotted and continuous curves refer to $W/\frac{Z^2 e^2 \ell^2}{32\pi}$ and $W'/\frac{Z^2 e^2 \ell^2}{32\pi}$,
respectively. In both instances $y=2$.
The singularity in $W'$ is shifted to the left toward a smaller $f$ for higher values of $y$.}
\end{figure}

Using (\ref{7}), condition $-1\leq \cos\theta'\leq 1$ becomes
\begin{equation}\label{costh'}
\ell_1\equiv\frac{R}{8\sigma p_{0}(n_{f}-\sigma)}\leq \ell\leq \frac{-R}{8\sigma p_{0}(n_{f}+\sigma)}\equiv \ell_2\,.
\end{equation}
The l.h.s. of (\ref{costh'}) is negative because $R<0$ and $n_f\geq \sigma$, therefore $\ell_1$ must be replaced by $0$. When $\ell_2$ is substituted in
(\ref{7}), we find $\cos\theta'\sim -R/(8m\sigma^2 y^{-})$ where $y^{-}=\sqrt{(1-6f^3 /(1+f)^3)/(1+f)} $ is the minimum value of $y$ for which $n_f$ is real.
$\cos\theta'$ has the maximum value $0.22$ at $f\sim 0.6$, it therefore follows that $\theta'\leq 0.43 \pi$.
$W$ and $W'$ are compared in Fig.\ref{fig:QGEOM1}.

\section{Conclusions}

We have derived the radiation characteristics of a charged particle subject to acceleration regimes described by (\ref{eq1}).
A process described by the  lowest order diagram of Fig.\ref{fig:Feynman2} is kinematically forbidden unless
the dispersion relations of anyone of the particles involved is altered either by a
medium, or by a gravitational field. The incoming particle "feels", in the case considered, a gravitational field given by (\ref{eq1}).
We have therefore calculated the index of
refraction due to MA which implies
the metric structure (\ref{eq1}) and can therefore be treated as a gravitational field. The calculation follows two distinct approaches.
The first one, more suitable for low-acceleration regimes, is based on the method discussed in \cite{caipap,singh,pap1} that leads to solutions of the covariant Dirac equation that are exact to $\mathcal{O}(\gamma_{\mu\nu})$.
In this approach the momentum remains finite, even in the case of a self-accelerating particle, but $N_f$
has singularities when $f=-2y^2/(1+y^2)$. The spectrum is typically $\propto \ell$ and the
infrared divergence of $W$ for $y\rightarrow 0$ is of no consequence. In addition, if $p\gg m$, the process is possible if $f<-2/3$
and radiation is in the forward direction.

In the second approach we take into account the physical singularity in
the Ricci scalar $R$ by using the second order differential equation that can be obtained from
the Dirac equation. This approach is better suited to study the high-acceleration regime.
The corresponding index of refraction $n_f$ has singularities when
$(1+f)(1+y^2) +6f^3/(1+f)^3 =1$. Radiation is possible in this case if $0<\ell \leq \ell_2$
and is contained in the cone of aperture $\theta' \leq 0.43\pi$.

In the effective geometry (\ref{eq1}), charged, accelerated fermions radiate according to (\ref{W}) and (\ref{WW}).
On the other hand fermions in "true" gravitational fields also emit radiation in the process represented by Fig.\ref{fig:Feynman2}\cite{PRD82}.
It is in fact the index of refraction that by altering the dispersion relations of the fermion,
and of any other particle in general, makes the process possible in vacuo. Note that unlike the case of
\v{C}erenkov radiation, it is not required that the photon
be emitted in a medium.

The radiation power laws differ according
to whether one considers the low-acceleration, or the high-acceleration regimes.
By comparing $W$ and $W'$ with
the power radiated by a fermion in a gravitational field \cite{PRD82}
\begin{equation}\label{GF}
W_{GF}= (Ze)^2 \frac{GM}{b}\frac{\ell^2}{y^2}\,,
\end{equation}
where $b$ is the impact parameter of the source, we find that in both instances
$GM/b$ is replaced by factors of $f$, while the dependence on $(\ell/y)^2$ remains the same.
Quantum mechanics is, of course, formulated in terms of particle momenta and one therefore expects a dependence on y in (\ref{GF}),
as well as in (\ref{W}) and (\ref{WW}).  Note, however that for $b\sim R$, the factor $GM/R$ is a characteristic of the source, while
$f$ itself depends on $m$ via $\mathcal{A}_m$ and changes essentially from particle to particle. For $GM/b\sim f$ the radiation emitted by
the charged fermion is such that $W_{GF}\simeq W$, while
$W'\simeq W_{GF}$ if
$GM/b\sim |-1+y\sqrt{1+f}/\sqrt{(1+f)(1+y^2)+6 f^3/(1+f)^3-1}|\sim 1$
at extreme accelerations, provided $f$ applies in each case to the same particle. It therefore follows, from a comparison
of $W$ and $W'$ with (\ref{GF}), that {\it Caianiello's model allows, in principle, the use of charged particles as
tools to distinguish inertial from gravitational fields locally}. This is an interesting feature of the model that is not permissible
in a purely classical theory \cite{DEWITT,ROHR}. The result is a direct consequence of the quantum violations
of the equivalence principle that MA introduces. These violations disappear when $\hbar\rightarrow 0$.

Because the role of the gravitational field is so essential in the process of Fig.\ref{fig:Feynman2}, it is expected
that the procedures outlined in this work and in \cite{PRD82,GAL} will be of use in extended theories of gravity \cite{CAPOZ}.

\end{document}